# ЭМИССИОННЫЕ АТОМНЫЕ СПЕКТРЫ.
# ИНДИВИДУАЛИЗИРОВАННЫЕ КОМПЬЮТЕРНЫЕ СИМУЛЯЦИИ ЛАБОРАТОРНЫХ РАБОТ.

А.Д. Заикин[1], А.А. Заикин[2]
Новосибирский государственный технический университет, г. Новосибирск, Россия[1]
независимый исследователь, г. Потсдам, Германия[2]

Представлен опыт создания и эксплуатации компьютерных симуляций лабораторных работ по атомной физике. Конкретная предметная область работы – эмиссионные атомные спектры. Базовый компонент каждой лабораторной работы – компьютерный симулятор спектрографа, формирующий на экране монитора линейчатые спектры атомов. В основе симулятора лежит преобразование длины волны излучения в RGB триплет. Градуировка спектрального прибора осуществляется по атомарному спектру ртути.

Набор параметров установки и условий эксперимента уникален для каждого студента, выполняющего лабораторную работу на симуляторе. Индивидуализированные параметры, хранящиеся в Google-таблицах, передаются в html-шаблон лабораторной работы посредством сервиса скриптов приложений Google Apps. Индивидуализация параметров симулятора стимулирует студента к самостоятельной работе.

**Ключевые слова:** спектрограф, лабораторная работа, симулятор, скрипт Google Apps.

## 1. Введение

В рамках стандартного университетского лабораторного практикума по физике существует значительное количество работ, которые можно классифицировать как «черный ящик». Исследуемое физическое явление скрыто от непосредственного наблюдения, контроль процесса осуществляется посредством электрических измерительных приборов. Регистрируя показания приборов при выполнении такой работы, отличить физическую реальность от её имитации затруднительно.

В [1] изложен опыт создания и эксплуатации компьютерных симуляций лабораторных работ по механике, электричеству и магнетизму – работы в этих разделах физики удовлетворяют описанным условиям. Используемые в них измерительные приборы – вольтметры и амперметры.

Компьютерные симуляторы таких лабораторных работ не требуют реалистичности экранных изображений. Схематичность представления лабораторной установки существенно упрощает процесс создания симулятора, а контролируемые преподавателем и индивидуальные для каждого студента параметры виртуальной лабораторной установки стимулируют его к самостоятельности при выполнении работы.

В противовес описанной выше ситуации во многих лабораторных работах используется сложное и дорогостоящее оборудование. При этом изучение курса общей физики не предполагает профессионального освоения студентом такого оборудования. Студент должен понять принцип его работы, произвести несложную настройку, освоить рутинную операцию измерения. Решив эти задачи, можно переходить к главному: измерению, обработке результатов измерения, выявлению закономерностей, сравнению результатов измерений с теоретическими моделями.

Трудозатраты на освоение студентом оборудования могут превышать таковые на освоение предметной области. Отсюда стремление использовать в учебном процессе оборудование с минимальным числом степеней свободы, настроенное и отъюстированное, работу которого можно описать словами «включил-измерил».

В ином случае возникает глубокая вовлеченность учебного персонала в процесс выполнения студентом лабораторной работы. Данное обстоятельство находит отражение в учебных материалах. Так, учебное пособие для выполнения лабораторной работы [2] содержит такой пассаж: «Внимание! Любые операции по включению или выключению компьютера, запуску и настройке программы, устранению технических неполадок при работе установки или компьютера выполняет сотрудник лаборатории!», [3] – «…попросить преподавателя или лаборанта отрегулировать прибор» », [4] – «Обычно установка уже готова к работе и не требует юстировки. О необходимости выполнения настройки аппаратуры спросить у преподавателя или лаборанта».

Все это напрямую относится к лабораторным работам по изучению эмиссионных и адсорбционных оптических спектров. Сложное и дорогое оптическое оборудование нуждается в точной настройке. При этом в настоящих время визуальные наблюдения спектральной картины практически вытеснены регистрацией спектра с помощью электронно-оптических преобразователей, а использование компьютерной техники для управления экспериментальной установкой, хранения, обработки и визуализации результатов скорее правило, чем исключение.

Поэтому с методической точки зрения кажется вполне допустимым и разумным использование компьютерных симуляторов спектральных работ в лабораторном практикуме по общей физике. Наглядность эксперимента от такой подмены сильно не пострадает.

С учетом выше сказанного, была сформулирована задача – оставаясь в рамках подхода, изложенного в [1], разработать индивидуализированные компьютерные эмуляторы лабораторных работ по теме оптические эмиссионные спектры. Достигнутые результаты представлены в данной работе, это три лабораторные работы, названия которых приведены в списке:

- «Определение постоянной Ридберга по эмиссионному спектру атомарного водорода»,
- «Измерение изотопического сдвига линий спектров водорода и дейтерия»,
- «Эмиссионный спектр водородоподобного иона».

## 2. Технологические аспекты разработки симуляторов

Компьютерные симуляторы лабораторных работ реализованы в виде кросплатформенного web-приложения. Приложение включает в себя следующие компоненты: стартовая web-страница, html-шаблоны лабораторных работ и соответствующие им электронные таблицы Google Sheets.

Стартовая web-страница предназначена для формирования запроса на выполнение конкретной лабораторной работы. Идентификационная информация: учебная группа, фамилия и пароль, полученный у преподавателя, вводится студентом в соответствующие поля. Нажатие кнопки «Выполнить» запускает PHP-скрипт, который загружает html-шаблон соответствующей лабораторной работы, содержащий схемы, рисунки, управляющие элементы и Java-скрипты, реализующие математическую модель исследуемого физического процесса.

Наряду с общими элементами шаблон содержит персональные данные студента (фамилию и номер группы) и индивидуализированные для него параметры лабораторной установки. Индивидуализированные данные содержатся в Google-таблице. В ней же ведется протоколирование работы студента. Такая таблица создается для каждой лабораторной работы.

Доступ к Google-таблице осуществляется посредством сценария на платформе Google Apps Script, опубликованного как web-приложение и имеющего уникальный URL. Сценарий Google Apps Script проводит верификацию введенных персональных данных. Если верификация успешна, то формируется строка индивидуальных параметров, возвращаемая в зашифрованном виде в шаблон лабораторной работы.

Некоторые индивидуализированные параметры являются скрытыми, они предназначены для функционирования математической модели физического процесса. Часть из них определяется студентом в процессе выполнения лабораторной работы.

Открытые параметры отображается на экране, включаются в протокол лабораторной работы и являются дополнительными идентификаторами студента. Обычно это известные параметры лабораторной установки.

Разработку компьютерных симуляторов лабораторных работ в области спектрального анализа предварим анализом работы соответствующих приборов.

### 3. Монохроматор УМ-2

Принципиальную схему значительной части спектральных приборов воплощает монохроматор УМ-2. Его общий вид приведен на Рис. 1. На оптической скамье располагается рейтор с источником света (на рисунке он справа). Свет от источника последовательно проходит фокусирующую линзу, щель коллиматора, объектив коллиматора, дисперсионную призму, установленную на поворотном столике, и выходную трубу, состоящую из проекционного объектива, щели и окуляра.

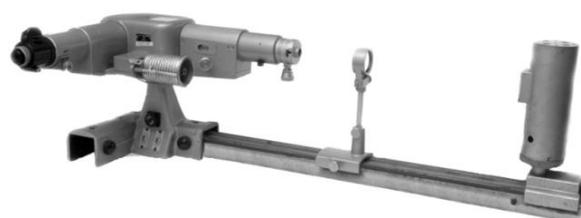

Рис. 1

Вращая барабан поворота призмы, Рис. 2, можно последовательно наблюдать в окуляре монохроматора отдельные спектральные линии. Указатель поворота барабана смещается по винтовой дорожке с нанесенными на нее делениями – это шкала поворотного механизма.

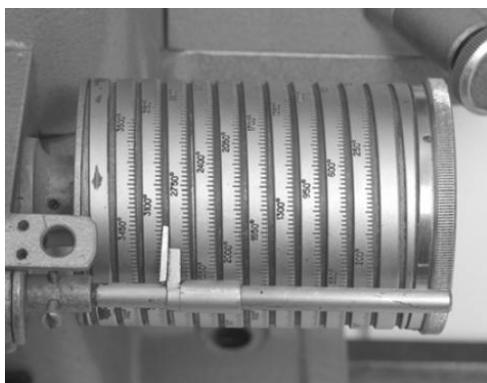

Рис. 2

Градусные деления, нанесенные на винтовую дорожку барабана монохроматора, носят относительный характер. Для того чтобы связать их с длинами волн эмиссионного спектра используется излучение с известными характеристиками. Чаще всего для градуировки спектрального прибора используется ртутная лампа низкого давления (например, дуговая ртутная спектральная кварцевая лампа типа ДРСк-125), являющаяся своего рода стандартом, с которым сравниваются все другие источники.

Возможности монохроматора, наглядно демонстрирующего принцип работы спектрального прибора, путем конструктивных изменений могут быть существенно расширены. Использование в качестве диспергирующего элемента дифракционной решетки значительно повышает разрешающую способность прибора.

Выходная щель монохроматора, располагающаяся в фокальной плоскости проекционного объектива, выделяет излучение, принадлежащее узкому интервалу длин волн.

Заменив щель протяженным приемником (например фотопластинкой), можно сформировал спектрограмму – набор цветных линий, располагающихся в соответствии с длиной волны. Не являясь единственным, такой способ визуализации спектра кажется наиболее естественным.

Получившийся спектральный прибор называется спектрографом. Для регистрации спектра в спектрографах применяются различные фотоэлектрические преобразователи, которые преобразуют энергию излучения в электрический сигнал. График зависимости энергии выходного излучения от длины волны также называют спектром.

## 4. Цифровая эмуляция спектрографа

В [5] описана цифровая эмуляция спектрографа, реализованная в информационной системе «Электронная структура атомов». Как указывают авторы, эта система, воссоздавая спектрограммы путем их эмуляции по базе спектральных данных атомных систем, позволяет решать задачи накопления и сравнения спектрограмм, определения методами спектрального анализа состава образцов, обучения специалистов и в целом ряде других случаев.

Базовый функционал любого эмулятора спектрографа – алгоритм преобразования длины волны излучения в триплет RGB, определяющий цвет спектральной линии на экране компьютерного монитора. Возможность представления цвета смесью красного, зеленого и синего (RGB цветовая модель) обусловлена особенностями строения глаза человека. Поскольку цвет – субъективное понятие, то такое преобразование не может не быть вариативным.

CIE (Commission Internationale Eclairage – Международная комиссия по освещению) разработала цветовое пространство CIE1931-RGB, в котором впервые были определены количественные связи между чистыми цветами и физиологическим восприятием цветов человеком [6]. Являясь по сути стандартом, CIE1931-RGB не практичен в расчетах, что стимулирует использование альтернативных подходов.

Так в [7] задаются семь опорных цветов: фиолетовый (380 нм), голубой (440 нм), бирюзовый (490 нм), зеленый (510 нм), желтый (580 нм), оранжевый (645 нм), красный (780 нм), а остальные цвета вычисляются методом линейной интерполяции.

Иной алгоритм преобразования предложил Дэн Брутон [8]. В этом алгоритме, а именно его будем использовать в данной работе, вклад каждой компоненты RGB триплета определяется функциями, графики которых приведены на Рис. 3, а спектр видимого света, построенный на основе данного алгоритма, показан на Рис. 4.

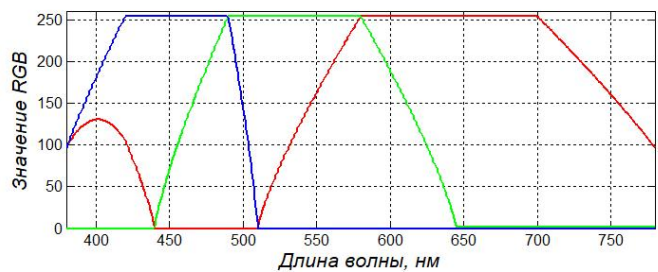

Рис. 3

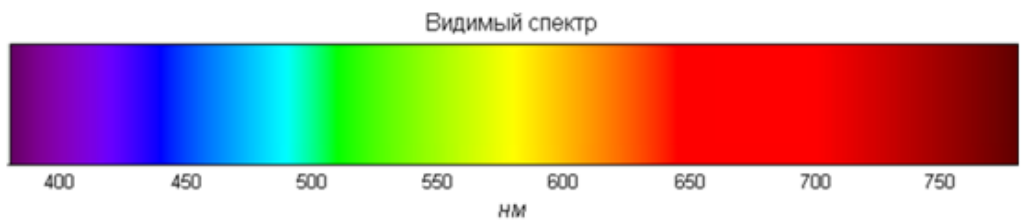

Рис. 4

Источником данных по атомарным спектрам ртути, водорода, дейтерия и некоторых других химических элементов, используемых в симуляторах, послужили работы [9,10,11]. Спектр молекулярного водорода взят в [12].

Эмулируя спектральный прибор, будем иметь в виду устройство и характеристики реального прибора – монохроматора УМ-2.

На Рис. 5 приведен фрагмент web-страницы, эмулирующей основные функции спектрографа

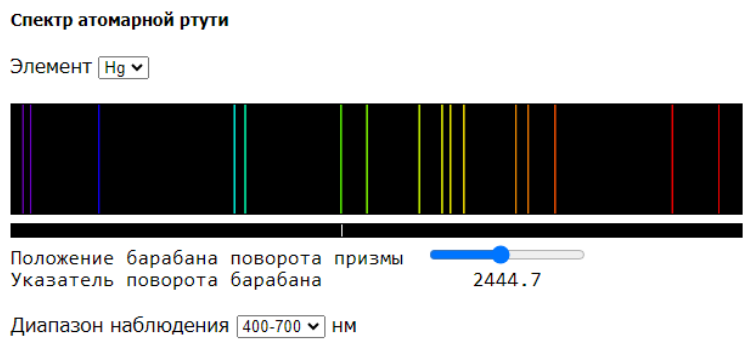

Рис. 5

На экран спектрографа выводятся спектральные линии оптического диапазона эмиссионного спектра выбранного химического элемента. Выбор осуществляется посредством выпадающего списка «Элемент».

Под экраном спектрографа располагается черное прямоугольное поле с вертикальным белым отрезком. Это указатель поворота барабана диспергирующей призмы. Его можно перемещать, добиваясь совпадения с определенной линией спектра, двигая ползунок (слайдер) положения барабана поворота.

Текстовое поле под ползунком отражает значение текущего градусного деления шкалы поворотного механизма. Выпадающий список «Диапазон наблюдения» позволяет масштабировать спектрограмму, наблюдая ее в разных разрешениях.

## 5. Градуировка спектрального прибора

Измерительным прибором монохроматор становится после градуировки. Обычно для градуировки используется спектр атомарной ртути.

Реальный спектр ртутной лампы носит сложный характер. В лампе наряду с атомами ртути могут присутствовать ее ионы и примеси. Технологически в трубку для облегчения зажигания вводится газ неон или аргон. Интенсивность линий спектра значительно различается, она определяется заселенностью соответствующих энергетических уровней. В силу ряда причин спектральные линии уширяются. На Рис. 5 приведен линейчатый спектр атома ртути в оптической части спектра. Далее под оптическим диапазоном будем понимать излучение с длинами волн 400-700 нм.

Спектр идеализирован, убраны частные особенности, всегда присущие конкретному эксперименту. Отобраны самые яркие линии, их интенсивность нормализована. Такой спектр можно использовать в качестве эталонного. Значения длин волн спектральных линий приведены в Таблице 1.

Градуировка монохроматора заключается в последовательном сопоставлении каждой спектральной линии, приведенной в таблице, и отсчета по шкале поворотного механизма (барабана).

Аналитическую форму градуировочной кривой можно получить, аппроксимируя методом наименьших квадратов результаты такого сопоставления полиномом второго порядка. Если $\varphi$ – отсчет по шкале барабана, то

Таблица 1  Спектр ртути

|    | Линия | Длина волны, нм |
|----|-------|-----------------|
| 1  | Фиолетовая левая | 404.656 |
| 2  | Фиолетовая правая | 407.783 |
| 3  | Синяя одинокая | 435.832 |
| 4  | Зелено-голубая левая | 491.606 |
| 5  | Зелено-голубая правая | 496.010 |
| 6  | Светло-зеленая левая | 535.400 |
| 7  | Светло-зеленая правая | 546.073 |
| 8  | Желто-зеленая | 567.600 |
| 9  | Жёлтая левая | 576.959 |
| 10 | Жёлтая средняя | 580.378 |
| 11 | Жёлтая правая | 585.925 |
| 12 | Оранжевая левая | 607.300 |
| 13 | Оранжевая правая | 612.327 |
| 14 | Красная одинокая | 623.400 |
| 15 | Красная левая | 671.634 |
| 16 | Красная правая | 690.746 |

$$\lambda = a + b\varphi + c\varphi^2 \qquad (1)$$

В дальнейшем, определив отсчет барабана, при котором наблюдается неизвестная спектральная линия, по градуировочной кривой можно определить соответствующую ей длину волны. Монохроматор становится измерительным прибором.

Параметры диспергирующей призмы являются уникальными характеристиками прибора. В рамках компьютерного эмулятора уникальность призмы, по сути, сводится к значениям трех коэффициентов полинома *a, b, c*.

Все представленные здесь лабораторные работы предполагают в качестве первого этапа градуировку монохроматора. Задав коэффициенты полинома в Google-таблице и сделав их уникальными для каждого студента, выполним поставленную задачу – индивидуализируем выполняемую лабораторную работу. Разумеется, коэффициенты *a, b, c* являются скрытыми параметрами, определяемыми в процессе выполнения работы.

## 6. Определение постоянной Ридберга по эмиссионному спектру атомарного водорода

Эмиссионный спектр атома водорода имеет линейчатый вид и распадается на серии (наборы линий). Спектральные линии, наблюдаемые в видимой части спектра, называются серией Бальмера. В нее входят четыре линии. Красная линия – длина волны 656.3 нм, обозначается как $H_\alpha$. Голубая – длина волны 486.1 нм, обозначается как $H_\beta$. Синяя – длина волны 434.1 нм, обозначается как $H_\gamma$. Фиолетовая – длина волны 410.2 нм, обозначается как $H_\delta$. Рис. 6 демонстрирует, как выглядит серия Бальмера на экране эмулятора.

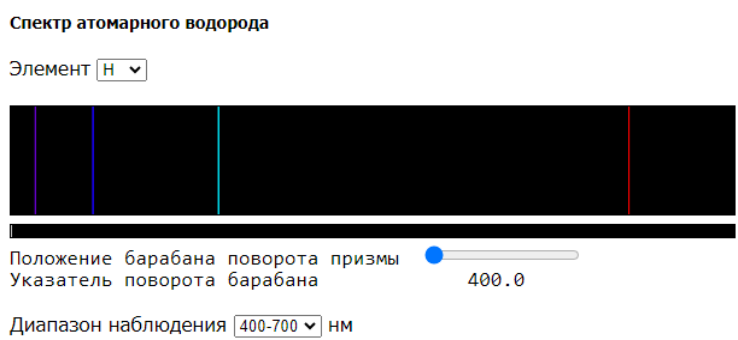

Рис. 6

Эмпирическая формула Ридберга определяет в спектре водорода двумя целыми числами наблюдаемые длины волн как в видимой области спектра, так и в инфракрасной и ультрафиолетовой

$$\frac{1}{\lambda} = R_\infty \left( \frac{1}{m^2} - \frac{1}{n^2} \right), \quad m = 1, 2, 3, \ldots, \quad n = m+1, m+2, \ldots \qquad (2)$$

Константа $R_\infty$ называется постоянной Ридберга. Являясь одной из наиболее точно измеренных фундаментальных физических постоянных, в настоящее время имеет численное значение $R_\infty = 10973731.568160 \; м^{-1}$.

Бор установил, что целые числа в формуле Ридберга (2) есть номера дискретных энергетических уровней электрона в атоме водорода. Номер уровня в современной терминологии – главное квантовое число. Излучение серии Бальмера реализуется при переходе электрона на второй энергетический уровень с вышележащих уровней.

Для наблюдаемых четырех линий серии Бальмера главные квантовые числа равны 3,4,5,6 в порядке убывания длины волны.

В теории атома Бора [13] постоянная Ридберга определяется через фундаментальные константы следующим образом $R_\infty = \dfrac{e^4 m_e}{8ch^3\varepsilon_0^2}$ .

С учетом вышесказанного экспериментальную проверку теории Бора и измерение постоянной Ридберга можно провести следующим образом. Измерив длины волн эмиссионного спектра атома водорода в видимой области и сопоставив их с главными квантовыми числами, можно построить график вида $\lambda^{-1} = f(n^{-2})$.

Если теория Бора верна, то график должен иметь вид прямой, Рис. 7.

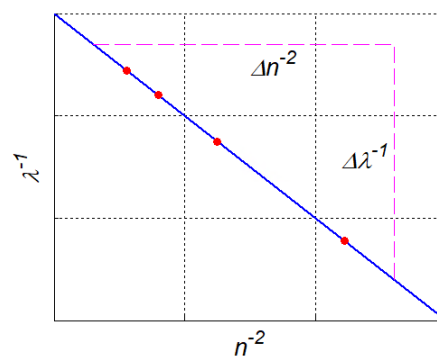

Рис. 7

Аппроксимировав результаты измерений линейной функцией, получим выражение вида

$$\lambda^{-1} = a - bn^{-2},$$

но по теории Бора значения коэффициента $a = R_\infty/4$, а $b = R_\infty$.

Угловой коэффициент прямой, величина которого равняется постоянной Ридберга, можно оценить и непосредственно по графику, разделив приращение функции на приращение аргумента, поскольку $b = |\Delta\lambda^{-1}|/\Delta n^{-2}$.

Лабораторная работа выполняется в несколько этапов. Первый этап – градуировка монохроматора. Зайдя на web-страницу симулятора и выбрав в выпадающем списке «Элемент» – *Hg*, студент формирует на экране линейчатый спектр атомарной ртути.

Перемещая указатель, определяет для каждой спектральной линии значение деления шкалы поворотного механизма. Строит градуировочный график и аппроксимирует результаты измерений полиномом второго порядка (1).

Второй этап. Выбрав в списке «Элемент» – *H*, студент формирует на экране линейчатый спектр атомарного водорода. Перемещая указатель, определяет для каждой спектральной линии значение деления шкалы поворотного механизма. Используя аппроксимационный полином (1), рассчитывает длину волны каждой спектральной линии.

Результаты измерений представляет в виде графика $\lambda^{-1} = f(n^{-2})$, и, рассчитав угловой коэффициент, получает постоянную Ридберга.

### 7. Измерение изотопического сдвига линий спектров водорода и дейтерия

Разрабатывая теорию атома, Бор исходил из предположения о том, что электрон вращается по круговой орбите вокруг бесконечно тяжелого, а значит, неподвижного ядра. Допущение вполне оправданное, о чем свидетельствует согласие предсказаний теории и эксперимента.

Вместе с тем, классическая механика утверждает, что в системе двух тел конечной массы каждое из них будет двигаться вокруг общего центра масс. Такое уточнение модели отразится на эмиссионном спектре. Наибольший интерес представляют различия, которые обнаруживаются в спектрах различных изотопов химических веществ – атомов, имеющих одинаковое зарядовое число, но различное массовое. Значим этот эффект прежде всего для легких атомов.

Ограничимся самым легким из них. Известны три изотопа водорода: протий, дейтерий и тритий, различающиеся количеством нейтронов в ядре. Под водородом обычно подразумевается именно протий. Ядро протия состоит из протона, дейтерия – из протона и нейтрона, трития – из протона и двух нейтронов. Эмиссионный спектр любого из этих изотопов согласно Бору определяется одинаково

$$\frac{1}{\lambda} = R_\infty \left( \frac{1}{m^2} - \frac{1}{n^2} \right) , \quad m = 1, 2, 3, \ldots , \quad n = m+1, m+2, \ldots , \quad R_\infty = \frac{e^4 m_e}{8 c h^3 \varepsilon_0^2} .$$

Индекс постоянной Ридберга показывает, что масса ядра полагалась бесконечно большой (неподвижное ядро).

Изложенный в [13] расчет, учитывающий движение ядра, приводит к тому, что в постоянную Ридберга вместо массы электрона войдет приведенная масса системы ядро – электрон.

$$\mu = \frac{m_e}{1 + m_e/M_я} .$$

Но это означает, что постоянная Ридберга, скорректированная с учетом движения ядра, будет различной для разных изотопов. Для протия и дейтерия постоянные Ридберга примут вид

$$R_H = \frac{R_\infty}{1 + m_e/M_H} , \qquad R_D = \frac{R_\infty}{1 + m_e/M_D} .$$

Различия масс ядер изотопов приводят к тому, что спектральные линии их эмиссионного спектра сдвигаются. Это явление называется изотопическим сдвигом.

Запишем разницу обратных длин волн излучения протия и дейтерия

$$\frac{1}{\lambda_D} - \frac{1}{\lambda_H} = \frac{R_\infty m_e (M_D - M_H)}{(M_D + m_e)(M_H + m_e)} \left( \frac{1}{m^2} - \frac{1}{n^2} \right).$$

Преобразуем это выражение

$$\frac{\lambda_H - \lambda_D}{\lambda_H} = \frac{m_e(M_D - M_H)}{M_D(M_H + m_e)} \frac{\lambda_D R_\infty}{(1 + m_e/M_D)} \left( \frac{1}{m^2} - \frac{1}{n^2} \right) = \frac{m_e(M_D - M_H)}{M_D(M_H + m_e)} \qquad (3)$$

Относительный изотопический сдвиг оказывается одинаков для всех спектральных линий.

Полагая, что $m_p$ – масса протона, а $m_n$ – масса нейтрона, запишем массы ядер изотопов

$$M_H = m_p, \quad M_D = m_p + m_n.$$

Если не учитывать различий в массах протона и нейтрона, $m_p \approx m_n$, и пренебречь вторым порядком малости отношения $m_e/m_p \ll 1$, то (3) можно преобразовать к виду

$$\frac{m_p}{m_e} \approx \frac{\lambda_H}{2(\lambda_H - \lambda_D)}. \qquad (4)$$

Таким образом, экспериментальное определение изотопического сдвига излучения водорода и дейтерия позволяет рассчитать отношение масс протона и электрона.

Приступив к выполнению лабораторной работы, студент по спектру атомарной ртути выполняет градуировку монохроматора. Затем проводит измерение значения деления шкалы поворотного механизма для спектральных линий $\alpha, \beta, \gamma, \delta$ атома водорода и дейтерия, и, используя аналитическую зависимость градуировочной кривой (1), определяет их длины волн.

Изотопический сдвиг каждой из четырех спектральных линий серии Бальмера позволяет согласно (4) рассчитать соотношение масс протона и электрона. Четыре независимых измерения позволяют построить доверительный интервал для искомой величины и сверить его с табличным значением.

### 8. Эмиссионный спектр водородоподобного иона

Теория атома Бора оказалась применима не только непосредственно к атому водорода, но и к водородоподобным ионам. Под водородоподобным ионом понимают ион, состоящий из ядра с зарядом $+Z|e|$ ($Z$ – зарядовое число) и одного электрона. Например: $He^+$, $Li^{2+}$, $Be^{3+}$, $B^{4+}$ и т.п.

В 1897 году в спектре излучения звезды Дзета Кормы была обнаружена спектральная серия очень похожая на серию Бальмера эмиссионного спектра водорода. Серия получила название в честь астронома Пикеринга.

В серии Пикеринга, Рис. 8, выделяются две группы чередующихся спектральных линий. Одна из них практически совпадает с линиями серии Бальмера водорода. Малые различия вполне объяснялись изотопическим эффектом.

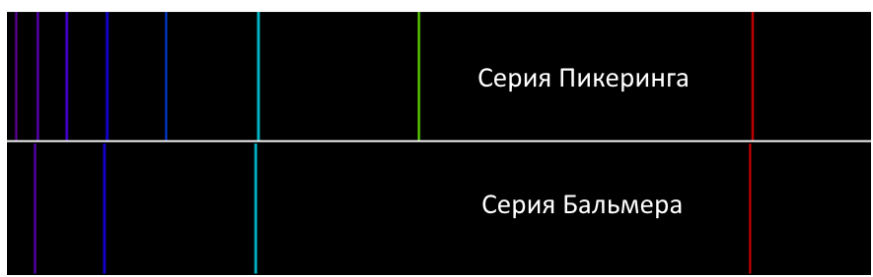

Рис. 8

Модифицированная формула Бальмера, допускающая как целые, так и полуцелые значения главного квантового числа, удовлетворительно описывала серию Пикеринга. Целые значения соответствовали линиям Бальмера, полуцелые линиям, располагающимся между ними.

Была высказана гипотеза об особом состоянии водорода в звездах. Однако Бор высказал альтернативную гипотезу о том, что эта серия соответствует водородоподобному иону.

Пренебрегая изотопическим эффектом, запишем формулу Ридберга для такого объекта

$$\frac{1}{\lambda} = Z^2 R_\infty \left( \frac{1}{m^2} - \frac{1}{n^2} \right), \quad m = 1,2,3,..., \quad n = m+1, m+2,..$$

Зарядовое число для атома водорода $Z=1$, а для водородоподобного иона равняется порядковому номеру в периодической таблице химических элементов.

Приравняв длину волны красной линии серии Бальмера $H_\alpha$ и линии спектра иона

$$Z^2 R_\infty \left( \frac{1}{m^2} - \frac{1}{n^2} \right) = R_\infty \left( \frac{1}{2^2} - \frac{1}{3^2} \right) ,$$

получим соотношения $2Z = m$ и $3Z = n$.

Проделаем то же самое для голубой линии $H_\beta$, тогда

$$Z^2 R_\infty \left( \frac{1}{m^2} - \frac{1}{(n+2)^2} \right) = R_\infty \left( \frac{1}{2^2} - \frac{1}{4^2} \right) .$$

Получаем, что $4Z = n+2$, но тогда $Z=2$, $m=4$, $n=6,7,8,...$.

Таким образом, Бор доказал, что серия Пикеринга – это излучение однократно ионизированного атома гелия, возникающее при переходе электрона на четвертый энергетический уровень с вышележащих, начиная с шестого. Излучение при переходе с пятого уровня на четвертый лежит в инфракрасной области и не попадает в оптический диапазон.

С учетом изотопического эффекта серия Пикеринга описывается формулой

$$\frac{1}{\lambda} = \frac{4R_\infty}{1+m_e/M_я}\left(\frac{1}{4^2}-\frac{1}{n^2}\right), \quad n=6,7,8,\ldots$$

Подтверждением правоты Бора стали лабораторные исследования искровых спектров гелия, в которых обнаружились спектральные линии серии Пикеринга. В искровых электрических разрядах достигается высокая степень ионизации, чего не наблюдается в дуговых спектрах.

С учетом вышесказанного, экспериментальные исследования эмиссионного спектра однократно ионизированного гелия позволяют определить массу атома. Для этого следует измерить длины волн серии Пикеринга, и, сопоставив их с главными квантовыми числами, построить график вида $\lambda^{-1}=f(n^{-2})$. Согласно теория атома Бора этот график должен быть линейным и количественно соответствовать выражению

$$\lambda^{-1}=a-bn^{-2}, \quad a=\frac{R_\infty}{4(1+m_e/M_я)}, \quad b=\frac{4R_\infty}{1+m_e/M_я}.$$

Угловой коэффициент прямой $b$ можно оценить непосредственно по графику, разделив приращение функции на приращение аргумента либо построив аппроксимационное уравнение. Тогда для углового коэффициента и массы атома справедливо

$$b=\frac{|\Delta\lambda^{-1}|}{\Delta n^{-2}}, \qquad M_я=\frac{bm_e}{4R_\infty-b}.$$

Если пренебречь разницей масс протона и нейтрона и малой массой электрона, то масса атома равняется массе ядра. Пусть химический элемент $X$ имеет $N$ нуклонов, тогда $M_X\approx Nm_p$. Зная массу атома, можно определить количество нуклонов, содержащихся в его ядре.

Лабораторная работа начинается с градуировки монохроматора. Затем, выбрав в списке «Элемент» – $He^+$, студент формирует на экране линейчатый спектр однократно ионизированного атома гелия. Перемещая указатель, определяет для каждой спектральной линии значение деления шкалы поворотного механизма, вслед затем рассчитывает длину волны каждой спектральной линии. Результаты измерений представляет в виде графика $\lambda^{-1}=f(n^{-2})$, и, рассчитав угловой коэффициент, определяет массу атома гелия и количество нуклонов в ядре.

## 9. Дополнительные задания

Лабораторные работы могут, по усмотрению преподавателя, содержать факультативные задания. Приведем доступные варианты.

Спектральные исследования позволяют определить химический состав пробы. Смесь атомов, маркированная в выпадающем списке «Элемент» символами XY, состоит из атомов трех химических элементов. Таблица 2 содержит используемую в работе выборку из 27 элементов. В ней же представлены аналитические (последние) спектральные линии этих элементов.

Таблица 2 Аналитические линии некоторых химических элементов в оптическом диапазоне

| № | Номер в таблице химических элементов | Символ химического элемента | Название | Длина волны аналитической линии, нм |
|---|---|---|---|---|
| 1 | 3 | Li | Литий | 670.78 |
| 2 | 6 | C | Углерод | 426.73 |
| 3 | 7 | N | Азот | 567.96 |
| 4 | 9 | F | Фтор | 685.60 |
| 5 | 10 | Ne | Неон | 640.22 |
| 6 | 12 | Mg | Магний | 518.36 |
| 7 | 17 | Cl | Хлор | 479.45 |
| 8 | 18 | Ar | Аргон | 696.54 |
| 9 | 20 | Ca | Кальций | 422.67 |
| 10 | 22 | Ti | Титан | 498.17 |
| 11 | 24 | Cr | Хром | 425.43 |
| 12 | 25 | Mn | Марганец | 403.08 |
| 13 | 30 | Zn | Цинк | 636.23 |
| 14 | 31 | Ga | Галлий | 417.21 |
| 15 | 35 | Br | Бром | 470.49 |
| 16 | 36 | Kr | Криптон | 587.09 |
| 17 | 41 | Nb | Ниобий | 405.89 |
| 18 | 47 | Ag | Серебро | 520.91 |
| 19 | 48 | Cd | Кадмий | 643.85 |
| 20 | 49 | In | Индий | 451.13 |
| 21 | 53 | I | Иод | 546.46 |
| 22 | 55 | Cs | Цезий | 455.54 |
| 23 | 56 | Ba | Барий | 553.56 |
| 24 | 57 | La | Лантан | 624.99 |
| 25 | 58 | Ce | Церий | 418.66 |
| 26 | 83 | Bi | Висмут | 472.26 |
| 27 | 88 | Ra | Радий | 482.59 |

Студент, выбрав вариант XY, получает на экране спектрограмму смеси. Определив длины волн этих линий, идентифицирует элементы, входящие в состав смеси, по таблице. Разумеется, набор химических элементов, входящих в состав смеси, индивидуализирован, и информация о нем содержится в Google-таблице.

Атомарный и молекулярный спектры излучения принципиально отличаются. Колоссальное количество спектральных линий молекулярного спектра, располагаясь весьма плотно, сливаются в полосы. Поэтому молекулярный спектр называется полосатым. Полосы разделяются узкими темными промежутками. Различить отдельные спектральные линии можно, лишь используя прибор с высоким спектральным разрешением.

В ознакомительных целях студент может сформировать на экране спектрографа полосатый спектр молекулярного водорода. Посмотреть его в диапазоне 400-700 нм, а затем в высоком разрешении, используя доступные поддиапазоны. Пример спектра молекулярного водорода в диапазоне 550-600 нм приведен на Рис. 9.

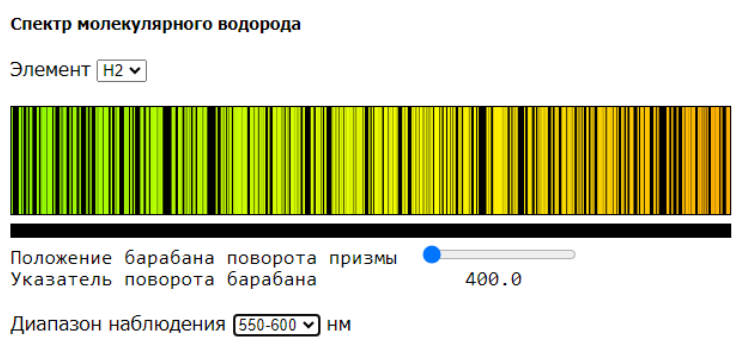

Рис. 9

Ознакомиться с работой эмулятора спектрографа в «гостевом» режиме можно, перейдя по [ссылке](). Такой режим ограниченной функциональности, не предполагающий индивидуализации характеристик монохроматора, достигается тем, что параметры диспергирующей призмы задаются следующими значениями *a=c=0, b=1*. Это приводит к тому, что градуировочная зависимость становится линейной, т.е. $\lambda = \varphi$. Фактически положение барабана поворота призмы в «гостевом» режиме градуировано в нанометрах.

## 10. Заключение

Создание компьютерного эмулятора спектрографа сделало возможным разработку виртуальных лабораторных работ по атомной физике. Объединяющее начало этих работ – эмиссионные атомные спектры, формируемые на экране компьютерного монитора. Спектрограммы оптической части спектра дополняются механизмом измерения относительного положения спектральных линий, а процедура градуировки по спектру атомарной ртути превращает спектрограф в измерительный прибор.

Линейчатые спектры атомов содержат в себе обширный массив информации о природе этих объектов. Измерив длины волн спектральных линий серии Бальмера атомарного водорода, можно провести экспериментальную проверку теории атома Бора и рассчитать постоянную Ридберга. Измерение изотопического сдвига спектральных линий водорода и дейтерия дает возможность сопоставить массы протона и электрона. По спектральным линиям серии Пикеринга (излучению однократно ионизированного атома гелия) можно определить массу ядра атома гелия и количество нуклонов в нем. Поэтому работа со спектрами даже в режиме компьютерного симулятора важное звено в процессе изучения физики.

Технологически лабораторные работы выстроены так, что параметры лабораторной установки, используемой студентом, индивидуализированы. Это достигается прежде всего вариабельностью параметров диспергирующей призмы монохроматора. Апробация представленных работ прошла в процессе изучения физики студентами инженерных специальностей технического университета.

Наработки и опыт, полученные в процессе создания компьютерных симуляторов будут полезны при разработке новых лабораторных работ. В этой связи перспективным кажется направление, связанное с молекулярными спектрами.